\documentstyle[12pt,psfig]{article}
\newcommand{\be}{\begin{equation}}
\newcommand{\ee}{\end{equation}}   

\newcommand{\bea}{\begin{eqnarray}}
\newcommand{\eea}{\end{eqnarray}}

\begin{document}
\xivpt
\title{Baryon Stopping and Strangeness Production in
 Ultra-Relativistic Heavy Ion Collisions\footnote{This Work is supported
by BMBF, DFG, GSI.}}

\author{{\bf L. Gerland, C. Spieles, M. Bleicher, H. St\"ocker}
\\[0.2cm]
{\small Institut f\"ur Theoretische Physik der J.W.Goethe Universit\"at}\\
{\small Robert-Mayer-Str. 8-10, D-60054 Frankfurt a.M., Germany}
\\[0.3cm]
{\bf C. Greiner}
\\[0.2cm]
{\small Institut f\"ur Theoretische Physik der J.~Liebig-Universit\"at}\\
{\small Heinrich-Buff-Ring 16, D-35392 Giessen, Germany}
\\[0.5cm]}

\maketitle
\begin{abstract}
The stopping behaviour of baryons in massive heavy ion
collisions ($\sqrt{s}\gg 10$AGeV) is investigated within different microscopic models.
At SPS-energies the predictions range from full stopping to virtually total transparency.
Experimental data are indicating strong stopping.\\
The initial baryo-chemical potentials and temperatures at collider energies
and their impact on the formation probability of strange 
baryon clusters and strange\-lets are discussed.
\end{abstract}

\newpage

\section{Motivation}
At CERN-SPS energies the gross features of the  baryon
dynamics are being studied extensively to date.
The current data are not understood theoretically. The model predictions for massive
systems range from strong stopping to transparency,
the production mechanisms for secondary particles
are not well known. For example, the increase of the $K/\pi$ ratios
and (anti-)hyperons from pp to AA-collisions measured
by many groups at the AGS and SPS, is still controversial.

There are different ideas to explain these observations, for example
colour ropes~\cite{hein} and rescattering of secondaries~\cite{matt}.
Another possibility to describe the experimental data is the 
Quark Gluon Plasma (QGP)~\cite{koch}, a deconfined phase where strangeness
should be produced abundantly.
Therefore it is important to measure directly which
model gives the right description of the reaction dynamics.

Strangelets contain a large number of delocalized quarks
($u...u,d...d,\\s...s$). They may serve as a proof for the transient existence of QGP.
It may not be possible to distinguish these multiquark droplets
from MEMOs~\cite{schaff} experimentally.

MEMOs are \underline{M}etastable \underline{E}xotic \underline
{M}ultistrange \underline{O}bjects (hyperon clusters),
which could be created by coalescence of hyperons. A QGP is not needed for such processes.  
Only in the interior of neutronstars or in relativistic heavy ion collisions
one can expect the simultaneous presence and phase-space-density of sufficiently many hyperons/strange quarks
to allow for the formation of multi-strange matter.

\begin{figure}[t]
\centerline{\hbox{\psfig{figure=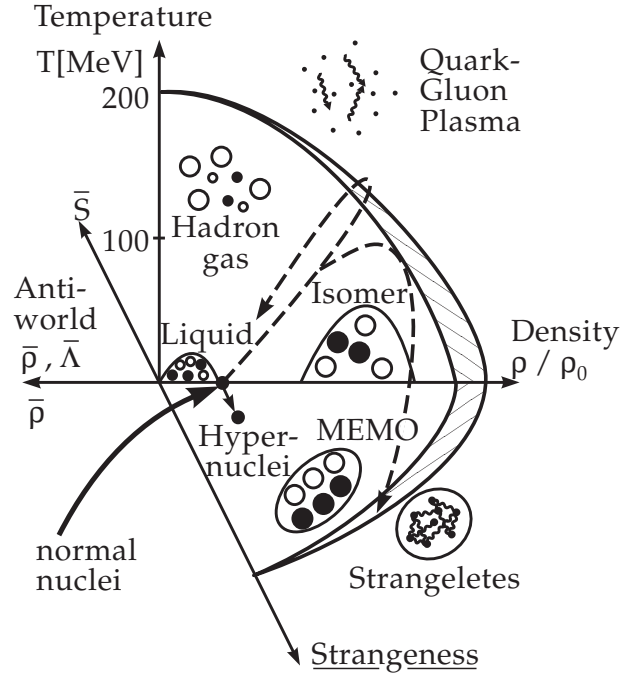,height=3.5in}}}
\caption{Sketch of the nuclear matter phase diagramm.}
\label{phased}
\vspace*{.5cm}
\end{figure}

Relativistic heavy ion collisions are the only tool to probe
hot and dense nuclear matter
under lab conditions.
 Fig.~\ref{phased} shows the phase diagram of nuclear matter and different paths
in the course of a heavy ion collision.
At finite baryon density, the strangeness is separated
from the antistrangeness due to associated production and evaporation~\cite{cg}.
This immediately drives the system off the $f_{s}=0$ plane into the
(anti-)strange sector, where parts of the system can cool down and form exotic objects
(as strangelets or MEMOs). 

We want to point out that
such states of matter can be created in
heavy ion collisions even at collider energies.
Let us first study shortly the main particle production mechanism in 
the used event generator FRITIOF~7.02~\cite{fritiof}.

\section{String-Fragmentation}
\begin{figure}[t]
\centerline{\hbox{\psfig{figure=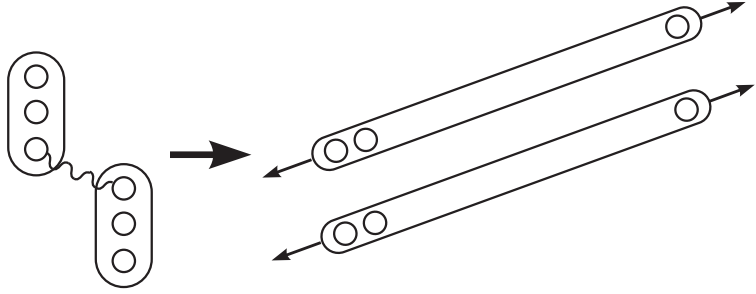,height=1.5in}}}
\caption{Creation of strings in nucleon-nucleon-scattering\protect\cite{mhof}.}
\label{eplemin}
\vspace*{.5cm}
\end{figure}

At higher energies excited nucleons (with masses higher than 2 GeV) cannot be described by resonances.
A frequently used model for the high energy excitation of the nucleons and
the subsequent particle production
 is the Lund string model~\cite{lstri},which is based on a 1+1 dimensional 
idealization of a colour flux tube. The excitation mechanism in Fritiof is the momentum transfer
between the constituent quarks as shown in Fig.~\ref{eplemin}, whereas for example the 
Dual Parton Model~\cite{ami} assume colour exchange as reason for the excitation.
The model is fitted to the production of hadrons in high energy $e^+e^-$-scattering,
so that there are no free parameters in nucleon-nucleon-collisions.
These strings decay into hadrons by a mechanism, which is for obvious reasons 
called "the tunneling process"~\cite{lstri}.
\begin{figure}[b]
\centerline{\hbox{\psfig{figure=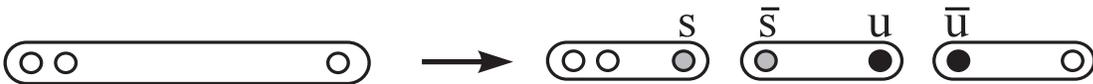,width=14.5cm}}}
\caption{Example of the decay of an baryonic string\protect\cite{mhof}.}
\label{fission}
\vspace*{.5cm}
\end{figure}
Motivated by the Schwinger-formalism \cite{schwinger}
for $e^{+}e^{-}$-pair-production in an infinite electric field,
we describe the production of $q \bar{q}$-pairs in the colour force-field
of a string with the formula:
\begin{equation}
\left| M\right|^{2} \propto {\rm exp}\left( -\frac{\pi m^2}{\kappa}\right)\quad.
\label{schwing}
\end{equation}
Here $|M|^2$ is the probability to produce a parton-antiparton pair with the mass m 
in a colour field with string-tension $\kappa$.
The string-tension of 1 GeV/fm leads to a suppression of the heavier strange quarks (s)
and diquarks (di), as compared to up (u) and down (d) quarks. 
The following input is used in our calculations :\\
u : d : s : di = 1 : 1 : 0.3 : 0.1,\\
corresponding to $m_s$ = 280 MeV.

Let us shortly summarize
results for the stopping power in nuclear collisions at collider regime
and the production of secondaries at the LHC.


\section{Stopping}
\subsection{What is stopping}
\begin{figure}[bht]
\centerline{\hbox{\psfig{figure=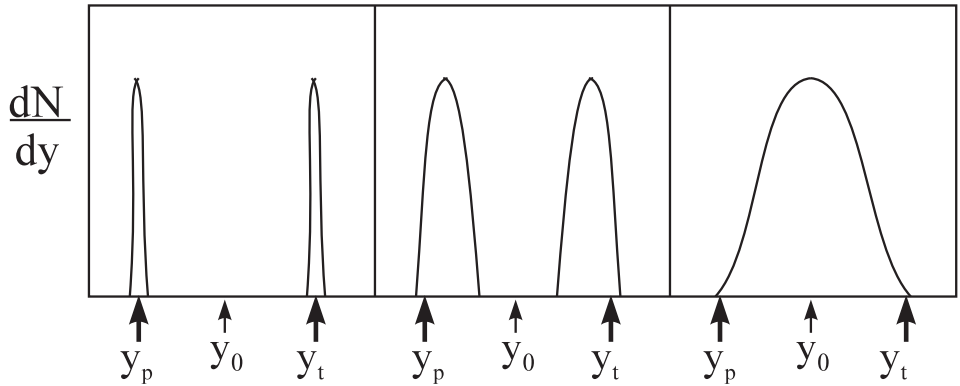,height=2in}}}
\caption{Idealized rapidity distributions\protect\cite{mhof}: Before the collision (a) and after the collision
with (b) transparancy and (c) stopping. $y_{p,t}$ are the rapidities of the projectile
and the target before the collision, $y_0$ is the mid rapidity. The distributions of
projectile and target are no $\delta$-functions before the collision because of the fermi momentum.}
\label{dndy}
\vspace*{.5cm}
\end{figure}
The rapidity distribution of baryons is used here to define stopping power in massive heavy ion collisions,
because the shape of a rapidity spectrum is Lorentz invariant. So this definition
does not depend on the chosen reference frame. 
As shown in Fig.~\ref{dndy},
a distribution of nucleons is designated as "stopped" after a nuclear collision,
if it has a maximum at midrapidity. The opposite scenario is called transparency.
This means that the projectile and target rapidity distributions may be smeared out,
compared to the initial distributions but still seperated by a nearly net-baryon-free midrapidity .
The longitudinal momentum  that is lost by projectile and target is used for the production 
of secondaries or transformed into the transverse direction. These two effects are not 
distinguished in this 
definition of stopping.
\begin{figure}[thb]
\centerline{\hbox{\psfig{figure=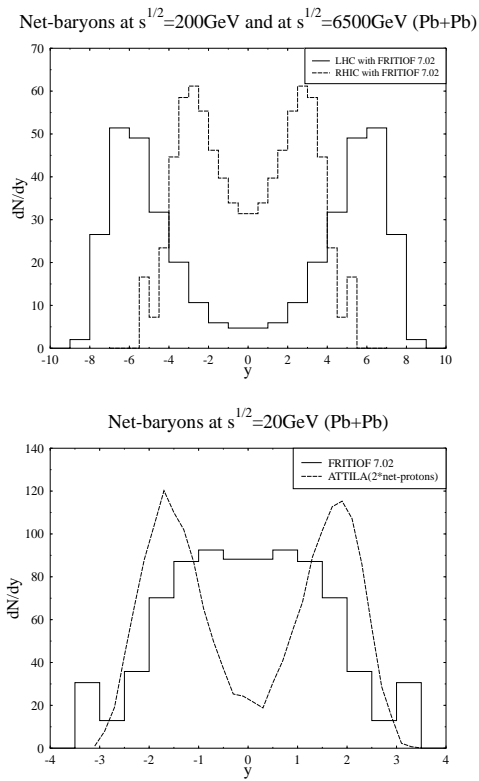,height=4.5in}}}
\caption{Net-baryon rapidity distribution of
very central Pb + Pb collisions at SPS, RHIC, LHC calculated with FRITIOF 7.02.
 The midrapidity region
 is even at LHC not net-baryon-free. For comparision the net-protons at SPS
calculated with ATTILA are also shown.}
\label{ger1}
\vspace*{.5cm}
\end{figure}  

\subsection{Stopping at collider energies}
\begin{figure}[hb]
\centerline{\hbox{\psfig{figure=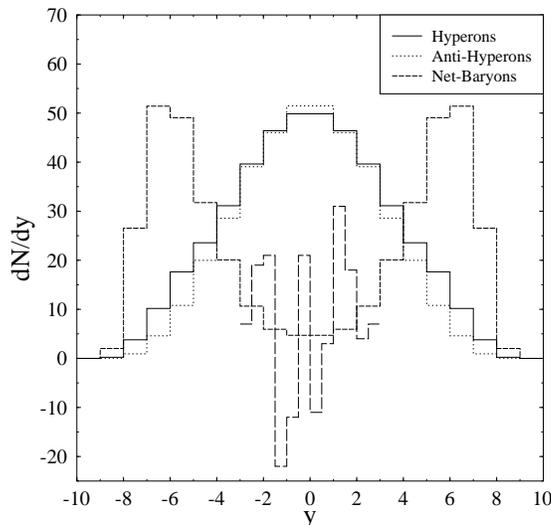,height=3.5in}}}
\vspace*{-2cm}
\caption{The (anti-)hyperon rapidity distribution of very
central Pb + Pb collisions at $\protect\sqrt{s}_{NN}=6.5$ TeV calculated with FRITIOF~7.02, and
mean net-baryon distribution at midrapidity compared
with the distribution of a single event.}
\label{ger2}
\vspace*{.5cm}
\end{figure}

Fig.~\ref{ger1} exhibits the baryon rapidity distribution as predicted by various 
models for heavy ion collisions. ATTILA~\cite{attila} 
and FRITIOF~1.7~\cite{frit} (not in the picture) show nearly a baryon-free midrapidity 
region already at SPS(CERN). 
These models are therefore ruled out by the new CERN data,
which rather support predictions based on the RQMD model~\cite{sorge}.
Also the new Lund model release FRITIOF~7.02 yields stopping at SPS! 
At RHIC FRITIOF~7.02 and RQMD~\cite{tomi} predict that the net baryon number
$A \gg 0$ at $y_{cm}$.
Furthermore, even in
very central collisions of lead on lead at $\sqrt{s}_{NN}=6.5$~TeV, there might be some
net-baryon density at midrapidity. 
This is shown in Fig.~\ref{ger2},
where the event-averaged rapidity densities of net-baryons, hyperons and
anti-hyperons are depicted for LHC, using FRITIOF~7.02.
If this non-perfect transparency turns out to be true, the finite baryo-chemical potential at 
midrapidity may have strong impact on the further evolution of the system.
As will be shown in section 5, expected yields of strangelets will be extremly sensitive to
the initial baryon-number of a Quark-Gluon-Plasma-phase.
\begin{figure}[bt]
\centerline{\hbox{\psfig{figure=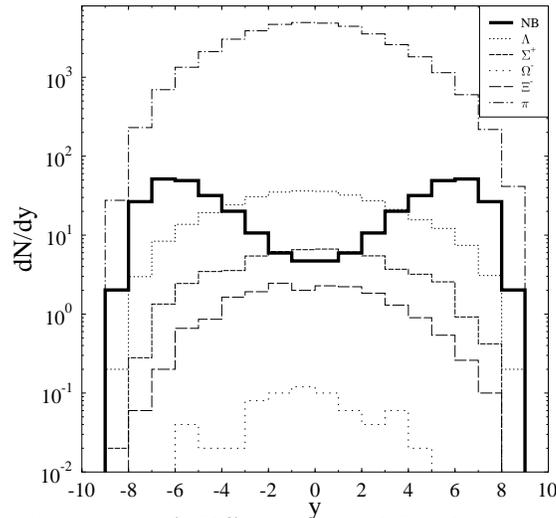,height=3.5in}}}
\vspace*{-2cm}
\caption{Rapidity distributions of different particles in very
central lead on lead collisions at $\protect\sqrt{s}_{NN}=6.5$ TeV calculated FRITIOF~7.02.
(NB denotes net baryons)}
\label{blei1}
\vspace*{.5cm}
\end{figure}

\section{Particle and Strangeness Production at LHC}

Fig.~\ref{blei1} shows the event-averaged rapidity densities of net-baryons, hyperons and
pions calculated with FRITIOF~7.02. Note that the strange to non-strange hadron 
ratios predicted by this model are the 
same for pp and AA collisions at 200 AGev/c (CERN-SPS) and that the strange particle
numbers for AA underpredict the data~\cite{antai}.
This deficient treatment of the collective effects in the model leads us
to take the numbers only as lower limits of the true strange particle   
yields at collider energies.

\begin{figure}[t]
\vspace*{-3cm}
\centerline{\hbox{\psfig{figure=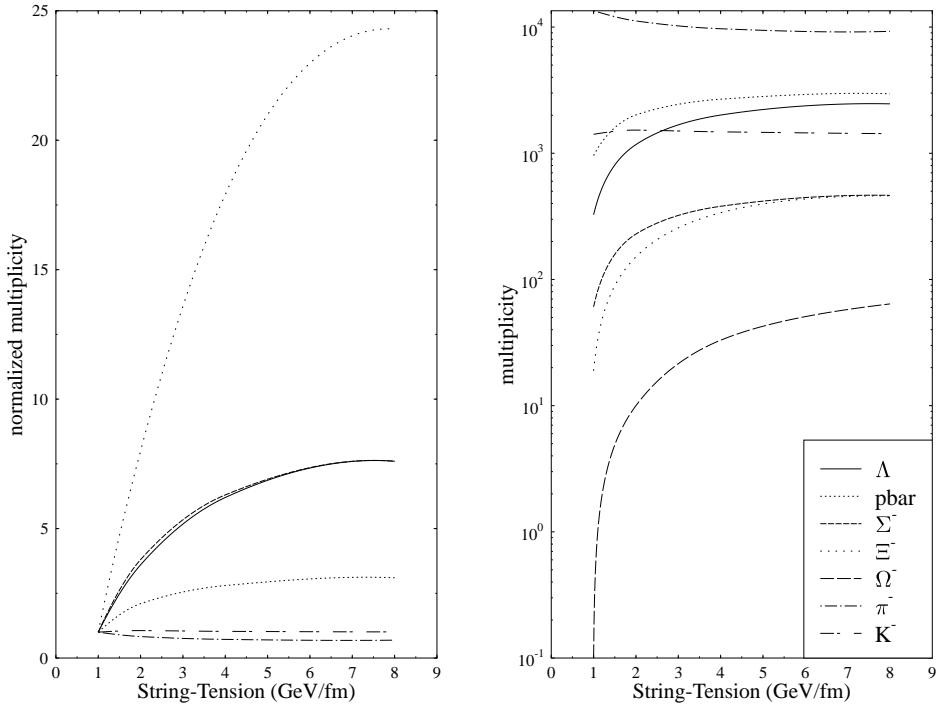,height=5in}}}

\caption{The multiplicities of different particles in very
central Pb + Pb collisions at LHC calculated with FRITIOF~7.02 as
function of the string-tension.}
\label{ger4}

\vspace*{.5cm}
\end{figure} 
Keep in mind that the microscopic models used here ignore possible
effects that could change significantly the number of produced strange particles
in heavy ion collisions, e.g. the string-string-interactions.
An enhanced string tension may effectively simulate
string-string interaction, as shown in Fig.~\ref{ger4}.
Here the multiplicities of different produced particles at LHC as function of
the string-tension $\kappa$ are depicted.
A higher string-tension, e.g. 2 GeV/fm yields  
the suppression factors :\\
u : d : s : di =  1 : 1 : 0.55 : 0.32.

\begin{figure}[thb]
\centerline{\hbox{\psfig{figure=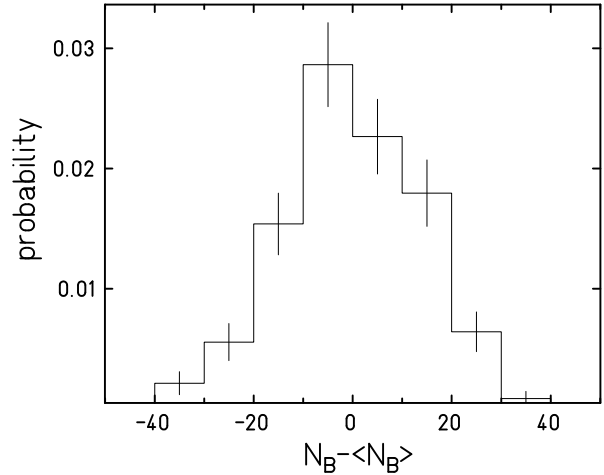,height=2.5in}}}
\caption{Probability distribution for net-baryon number fluctuations at mid-rapidity within bins of one unit
of rapidity width, calculated with FRITIOF 7.02 for the Pb + Pb system at $\protect\sqrt{s}_{NN} = 6.5$ TeV}
\label{ger3}
\vspace*{.5cm}
\end{figure}
To estimate the fluctuations in heavy-ion collisions at the LHC,
let us consider the above Pb+Pb calculation performed at $\sqrt{s}_{NN}=6.5$TeV. 
The probability distribution for non zero net-baryon number 
fluctuations at midrapidity is plotted in Fig.~\ref{ger3} within bins of one unit
 of rapidity width. This probability is defined for single events, and shows
rapidity density deviations from the average value $<dN/dy>$. 
Bins between $-3<y<3$ have been taken into account. The probability
for fluctuations $N_B-<N_B>$ being larger than $\pm 20$ is about 15 \%. The asymmetry of the 
histogram results from the fact that fluctuations may be different for positive 
and negative deviations around a non zero average rapidity density. 
This aspect is visible clearly in the $dN/dy$ of a randomly   
selected single event as drawn (for the mid-rapidity region) in
Fig.~\ref{ger2}. Now let us estimate the thermodynamic conditions, one can expect
$locally$ in a single event.
If each pion carries about 3.6 units of entropy (which is true for massless
bosons), the entropy per baryon content in the fireball is\\
\begin{equation}
\frac{S}{A_B} \, \approx \, 3.6 \, \frac{dN_{\pi}/dy}{dN_{B}/dy}
\,\,\, .
\end{equation}
This leads to a $S/A$-value of about 500 if we set $dN_{B}/dy=30$,
which seems to be possible in single events as discussed above.

We assume that the conditions estimated above within a purely had\-ronic model are present also
in the case of QGP creation.
If the plasma is equilibrated, the ratio of the
quarkchemical potential and the temperature $|\mu | /T$ is directly related
to the entropy per baryon number via
\begin{equation}
\left(\frac{S}{|A_B|}\right)^{QGP} \, \approx \, \frac{37}{15} \pi^2 \,
\left(\frac{|\mu |}{T}\right)^{-1}
\,\,\, .
\end{equation}
Accordingly the ratio then varies between 0.05 to 0.1.
In the following we adopt the model of ref.~\cite{PRD91}
for the dynamical creation of
strangelets out of QGP, and apply it to the hitherto unexplored collider
regime, assuming $\mu/T\stackrel{<}{\sim} 0.1$~\cite{spieles}.


\section{Strangelet Distillation}
\begin{figure}[thb]
\centerline{\hbox{\psfig{figure=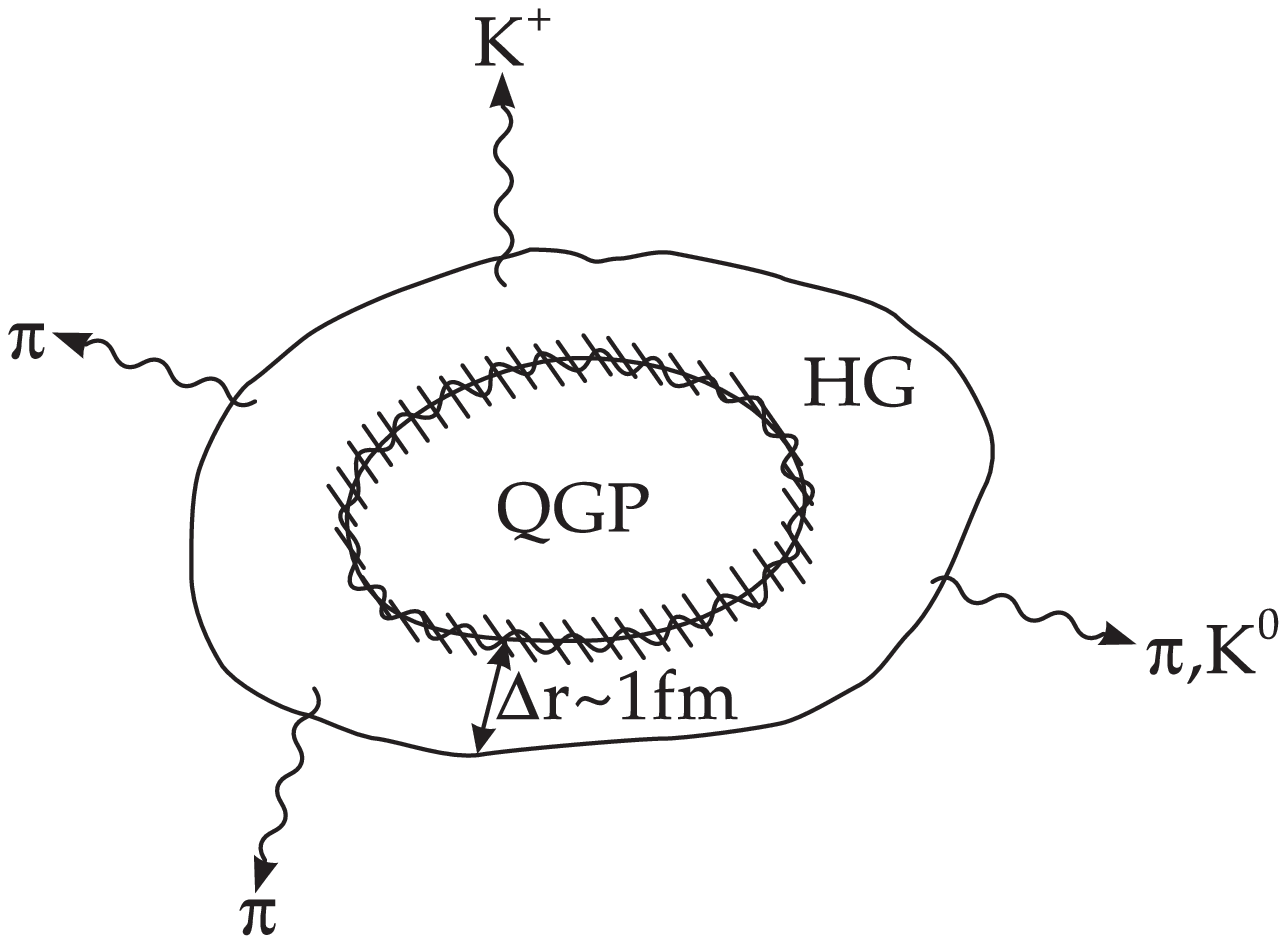,height=2.5in}}}
\caption{Hadron gas surrounds the QGP at the phase transition.
Particles evaporate from the hadronic region.
New hadrons emerge out of the plasma by hadronization.}
\label{qgp}
\vspace*{.5cm}
\end{figure}

Consider a hadronizing QGP-droplet with net-strangeness zero surround\-ed by a layer of hadron gas which 
continuously evaporates hadrons (they undergo the freeze-out).
Assume the two phases to be in perfect mechanical, chemical and thermal equilibrium.
Now, rapid kaon emission leads to a
finite {\em net} strangeness of the expanding
system \cite{PRD91}. As shown in Fig.~\ref{qgp},
this results in
an enhancement of the $s$-quark abundance in the quark phase.
Prompt kaon (and, of course, also pion) emission cools
the quark phase, which then may condense into metastable or stable
droplets of strange quark matter.

At collider energies, the midrapidity region is expected to be characterized 
by rather low net-baryon densities even when taking fluctuations into account. 
How can one expect to create stable strangelets
with baryon density of  $\rho > \rho_0$? 
\begin{figure}[bth]
\centerline{\hbox{\psfig{figure=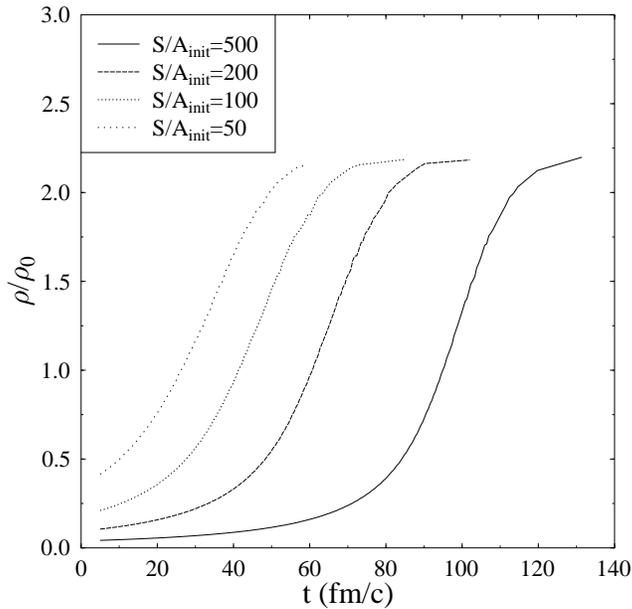,height=4in}}}
\vspace*{-1cm}
\caption{Time evolution of the net baryon density of a QGP droplet.
The initial conditions are
$f_s^{\rm init}=0$ and $A_{\rm B}^{\rm init}=30$.
The bag constant is $B^{1/4}=160$~MeV.}
\label{tdens}
\vspace*{-.5cm}
\end{figure}   

Fig.~\ref{tdens} illustrates
the increase of baryon density in the plasma droplet
as an inherent feature of the dynamics of the phase
transition.
This result originates from
 the fact that the baryon number in the
quark-gluon phase is carried by quarks with $m_{\rm q}\ll T_{\rm C}$, while
the baryon density in the hadron phase is suppressed by a Boltzmann factor
$\exp (-m_{\rm baryon}/T_{\rm C})$ ($m_{\rm baryon}\gg T_{\rm C}$).
A very tiny excess of initial net-baryon number will suffice to generate regions of
very high density $\rho_{\rm B}>\rho_0$! The very low initial $\mu/T$
corresponds to high values of the initial specific entropy.

\begin{figure}[th]
\centerline{\hbox{\psfig{figure=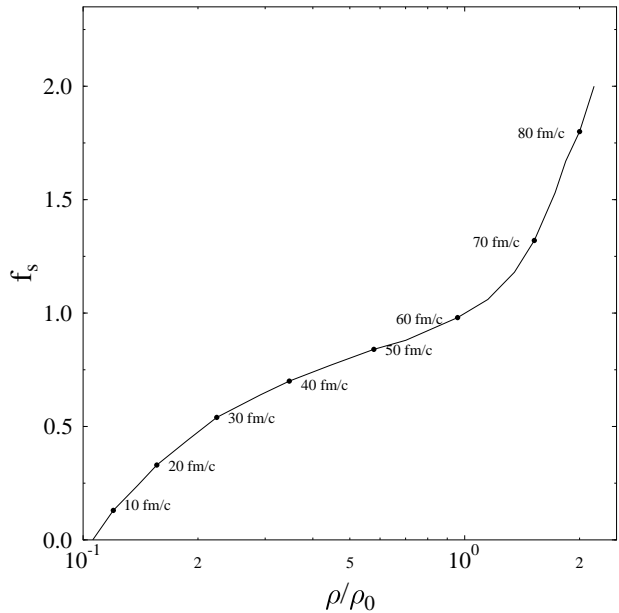,height=4in}}}
\vspace*{-1cm}
\caption{Evolution of a QGP droplet with baryon number $A_{\rm B}^{\rm
init}=30$ for $S/A^{\rm init}=200$ and
$f_s^{\rm init}=0$. The bag constant is $B^{1/4}=160$~MeV. Shown is the
baryon density and the corresponding strangeness fraction.}
\label{fsrho}
\vspace*{-.5cm}
\end{figure}

Fig.~\ref{fsrho} shows the evolution of the two-phase system for $S/A^{\rm
init}=200$, $f_s^{\rm init}=0$ and for a bag constant $B^{1/4}=160$~MeV 
in the plane of the strangeness fraction vs. the baryon density.
The baryon density increases by more than one order of magnitude!
Correspondingly, the chemical potential rises as drastically
during the evolution, namely from $\mu^i=16$~MeV to $\mu^f>200$~MeV.

The strangeness
separation mechanism~\cite{cg} drives the chemical potential of the strange quarks
from $\mu^i_s=0$ up to $\mu^f_s\approx 400$~MeV.
Thus, the thermodynamical and
chemical properties during the time evolution
differ drastically from
 the initial values! 
Low initial chemical potentials do not hinder the creation
of strangelets with high $\mu$.
However, this result depends crucially on the bag parameter and may change, if
finite size corrections are included.

\section{Conclusion and Summary}
\begin{itemize}
\item Stopping at SPS 
\end{itemize}
Early models that claim no stopping at SPS are ruled out by the new CERN data.
Models which show the right stopping behaviour at this energy
still predict a finite net-baryon number at RHIC ($A\gg0$) and LHC ($A \neq 0$)
 at midrapidity.
\begin{itemize}
\item Fluctuations of net-baryons and net-strangeness in single events
\end{itemize}
At collider energies, the physics of single events
may differ extremely from an event-average.
 Because of these fluctuations interesting physics can be expected, e.g.
 intermittency fluctuation, $\pi$-droplets, clusters and the above discussed exotic particles:
(anti-)strangelets (quark-droplets with $S<0$ and $A>1$) and (anti-)MEMOs
(hyperonic clusters) might be created.


\begin{thebibliography}{99}

\bibitem{hein}
H. Sorge, M. Berenguer, H. St\"ocker, W. Greiner,
Phys. Lett. B289 (92) 6

\bibitem{matt}
R. Mattiello, H. Sorge, H. St\"ocker, W. Greiner,
Phys. Rev. Lett. 63, (89) 1459

\bibitem{koch}
P. Koch, B. M\"uller, J. Rafelski,
Phys. Rep. 142 (1986) 167

\bibitem{schaff}
J. Schaffner, C. Greiner, H. St\"ocker,
Phys. Rev. C 46, (1992) 322

\bibitem{cg}C. Greiner, P. Koch and H. St\"ocker,
    Phys. Rev. Lett. {\bf 58}, 1825 (1987);
    C. Greiner, D.~H. Rischke, H. St\"ocker and
    P. Koch, Phys. Rev. D{\bf 38}, 2797 (1988) 

\bibitem{fritiof}
H. Pi, {\it Comput. Phys. Comm.} {\bf 71} (1992) 173-192

\bibitem{lstri} B. Andersson, G. Gustafson, G. Ingelman and T. Sj\"ostrand,
   {\it Phys. Rep.} {\bf 97} (1983) 31.

\bibitem{mhof}
M. Hofmann, diploma thesis, Johann-Wolfgang-Goethe Universit\"at Frankfurt am Main (1993)

\bibitem{ami}
A. Capella, U. Sukhatme, C. I. Tan and J. Tranh Van,
Phys. Rep. 236

\bibitem{schwinger}
J. Schwinger, Phys. Rev. {\bf 82} (1951) 664

\bibitem{attila}
M. Gyulassy, private communication

\bibitem{frit}
B. Nilsson-Almqvist and E . Stenlund,
   {\it Comput. Phys. Comm.} {\bf 43} (1987) 387

\bibitem{sorge}
H. Sorge, H. St\"ocker, W. Greiner,
Ann. Phys. (USA) {\bf 192} (1989) 266; Nucl. Phys. A{\bf 498} (1989) 567c; Z. f. Phys. C{\bf 47} (1990) 629

\bibitem{tomi}
Th. Sch\"onfeld, H. St\"ocker, W. Greiner, H. Sorge,
Mod. Phys. Lett. A8 (1993) 2631

\bibitem{antai}
A. Tai, B. Andersson and Ben-Hao Sa, Proc. of the Int. Conf. on Strangeness in Hadronic
Matter, S'95, Tucson, AZ, USA, AIP Press, Woodbury, NY (1995) p.335

\bibitem{PRD91}
C. Greiner and H. St\"ocker,
Phys. Rev. D{\bf 44}, 3517 (1991)

\bibitem{spieles}
C.~Spieles, L.~Gerland, H.~St\"ocker, C.~Greiner,
C.~Kuhn, J.~P.~Coffin,
Phys. Rev. Lett. 76 (1996) 1776

\end{thebibliography}
\end{document}